\documentclass{article}
\usepackage{hiph-art}
\usepackage{graphicx}
\def\rightmark{Semi-exclusive processes, grey track production and hadronization}
\def\leftmark{C. Ciofi degli Atti}
\title{Space-Time Evolution of Hadronization in DIS: \\
Semi-Exclusive Processes and Grey Track Production\footnote{Talk given at the Workshop 
 on
{\it In-Medium Hadron Physics},
Giessen, Nov. 11-13, 2004}} 
\authors{ 
{Claudio Ciofi degli Atti
}\\[2.812mm]
{\normalsize
Department of Physics, University of Perugia \&
INFN, Sezione di Perugia,\\
06122 Perugia, Italy\\[0.2ex] 
}} 
\abstract{Semi-exclusive processes and grey track production in DIS off nuclear targets
are analyzed as possible tools to investigate the space time evolution of hadronization.}

 
\makeindex
\begin{document}
\maketitle

\section{Introduction}\label{intro}
Nuclear targets serve as a natural and unique analyzer of the space-time
development of strong interactions at high energies. Due
to Lorentz time dilation, projectile partons may keep their coherence for
some time, but once they become incoherent  the cross section of final
state interaction (FSI) increases. One needs observables sensitive to
such modifications of FSI.
One way for experimental testing of theoretical models is the measurement of
the nuclear modification factor for inclusive production of leading
hadrons \cite{knp}, and  
various theoretical approaches have been advocated \cite{knph,ww,amp,giessen}
to explain recent data from HERMES experiment at HERA \cite{hermes}.
The production of hadrons with large fraction $z_h$ of the initial parton
momentum is a rare, nearly exclusive process, which has quite a specific
time development. In the main bulk of events the jet energy is shared by
many hadrons. It is a difficult task to find observables sensitive to the
time development of hadronization in this case, and additional, complementary processes
which could be sensitive to the space-time evolution of hadronization should be 
investigated. It is the aim of this paper to review two approaches which have been 
recently pursued along this line, namely the  Semi- Exclusive DIS (SEDIS)
 processes  $A(e,e'B)X$  \cite{ck,ckk}, where  
$B$ is a  detected recoiling 
heavy fragment (e.g a nucleus  $A-1$) and    $X$ the 
 unobserved jets from hadronization,  and the  grey track  production in DIS off nuclei
   \cite{grey}.
    \begin{figure}[hc]
\vspace*{-0.2cm}
\centerline{\includegraphics[height=4.0cm,width=6.0cm]{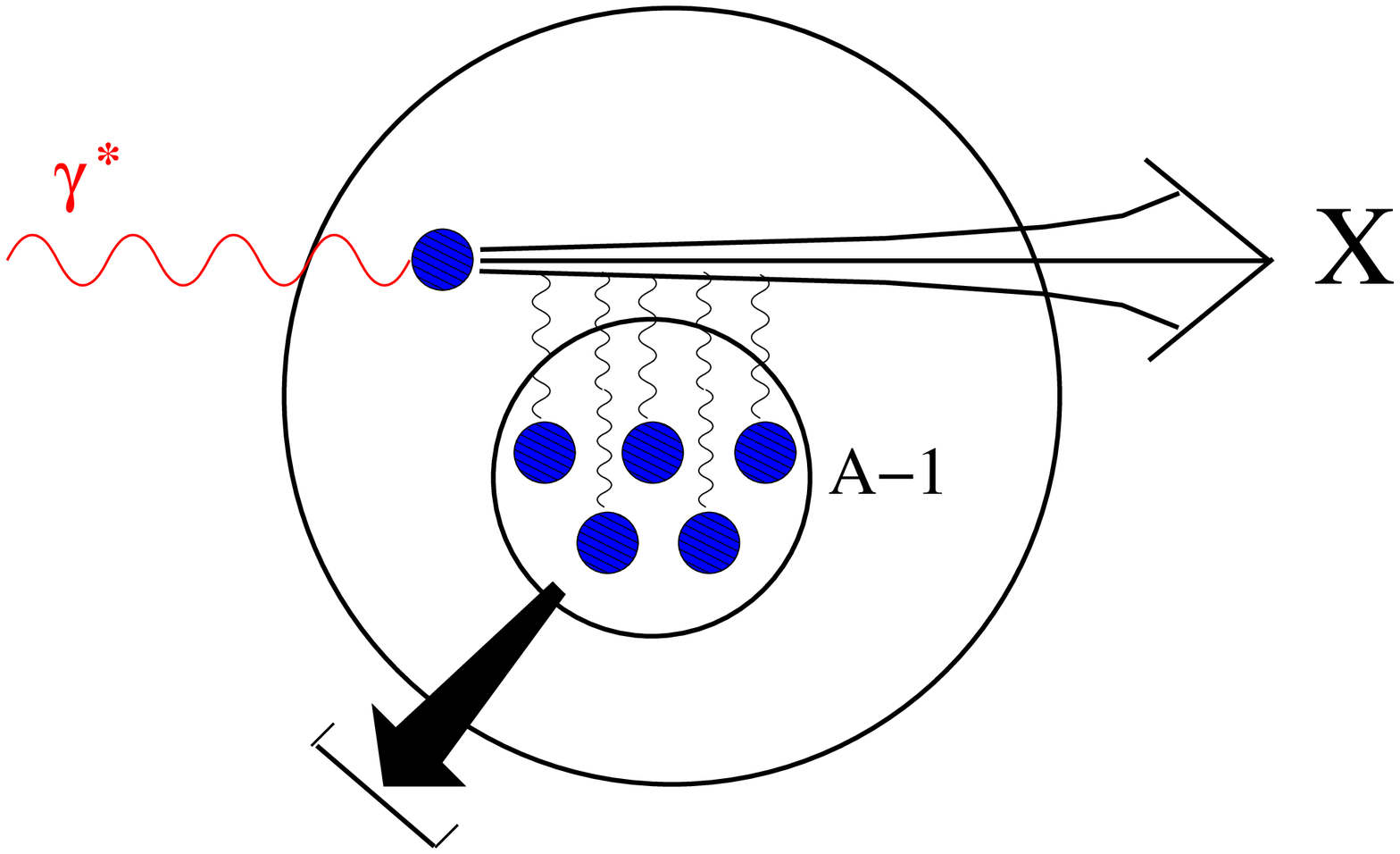}
\hspace {0.5cm}\includegraphics[height=4.0cm,width=6.0cm]{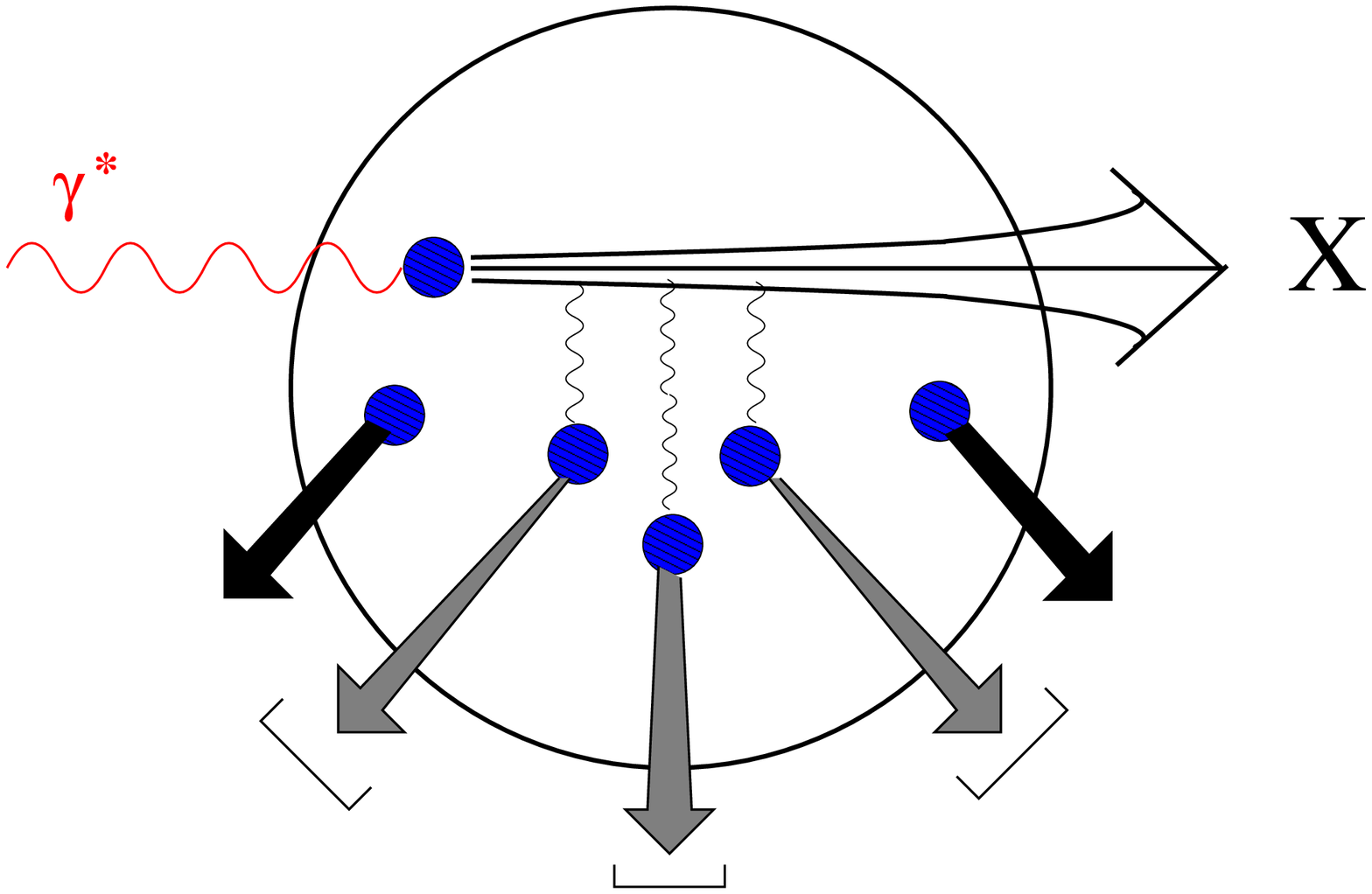}}
\vspace*{-0.2cm}
\caption[]{Cartoons of the process $A(e,e'B)X$ (with $B=A-1$) (left), and the grey track 
production (right) in DIS. {\it Left}:
 after $\gamma*$ interaction with a quark, the nucleon debris
interacts with the spectator nucleus $A-1$, which coherently recoils and is detected in coincidence with
the scattered lepton. 
{\it Right}
the nucleon debris breaks apart the spectator nucleus $A-1$, and nucleons with momentum
 $200-600 \,\,MeV/c$
are emitted and detected as grey tracks.}
\label{fig1}
\end{figure} 

   Unlike the rare process of leading hadron production, both processes,
   which are depicted in Fig. 1,
depend upon the   bulk  of the FSI of the nucleon debris with the 
nuclear medium,  and turn out to be very sensitive to the 
hadronization mechanism.  Their theoretical treatment 
 within a Glauber-like
 approach of FSI 
  requires an {\it effective time-dependent nucleon debris-nucleon cross
 section};  a recent model of the latter \cite{ck} 
  will be reviewed  in the next Section and its use in the theoretical description of the two processes will be presented in Sections
  3 and 4.

\section{The effective debris-nucleon cross section}\label{debris}

The effective debris-nucleon cross section obtained in \cite{ck}, is 
based upon   the 
hadronization model which combines the
soft part of the hadronization dynamics, treated  by means of the color string model,
with the hard part one,  described within perturbative QCD.
In this model, which is inspired  by Ref.  \cite{knp}, 
the formation of the final hadrons occurs
during and after the
propagation of the created  debris through the nucleus,
  with a sequence of soft and hard production  
  processes. Introducing a mass scale $\lambda=0.65\,\, GeV$ \cite{scale}, soft 
  production  occurs at  $Q< \lambda$ and  hard production  at 
   $Q> \lambda$; in
the former case $npQCD$ is taken care of by the color string model,
whereas in the latter case 
pQCD is described  within the gluon radiation  model \cite{knp}.
The details of the approach are given in Ref. \cite{ck}, here only the basic elements
of the model  will be recalled. The ingredients of the treatment of hadron (mostly mesons 
$({\bf M})$) formation
from the string $({\bf S})$ decay, are the following ones:
\begin{itemize}
 \item the probability $W(t)$ for a string to 
create no quark pairs since its origin;
 \item the time dependent length 
of the string $L(t)$ with  $L_{max}={m_{qq}}/{\kappa}$, where 
 $m_{qq}$ is the mass of the  "diquark"
 and $\kappa \simeq 1 \,GeV fm^{-1}$  the string tension;
 \item  the creation of a baryon ({\bf B}) and a shorter string
 after the  first breaking of the latter  (within
  $\Delta t \simeq 1\,fm$), and the sequence of  decays according to the scheme
  \begin{center}
   $\bf S$ $\Rightarrow$
  {\bf B}+$\bf S$ $\Rightarrow$
  {\bf B}+  $\bf S$+  {\bf M}
 $\Rightarrow$ {\bf B}+  $\bf S$+
{\bf 2} {\bf M}  +...
 \end{center}
 leading to the following multiplicity of mesons
 \begin{equation}
n_{M}(t)=
\frac{{\rm ln}(1+t/\Delta t)}
{{\rm ln}2}\ ,
\label{50}
\end{equation}
with $\Delta t \simeq 1 fm$.
 \end{itemize} 
 The gluon radiation mechanism is governed by:
 \begin{itemize}
 \item  the coherence time
  \begin{equation}
 t_c =\frac{2\,E_q\,\alpha\,(1-\alpha)}
{k_T^2}\ ,
\nonumber
 \end{equation}
which is the    time which elapses from the creation of the leading quark and the emission
of the  gluon which lost coherence with the color field of the quark. Here
   $\alpha$, $k_T=|{\bf k}_T|$, and
$E_q$ are the
fraction of the quark light-cone momentum carried by the radiated quantum, its 
 transverse momentum, and the quark energy, 
respectively;
  \item  the mean number of radiated gluons, given by
  \begin{equation}
 n_G(t)=
\int\limits_{\lambda^2}^{Q^2} dk_T^2
\int\limits_{k_T/E_q}^1  d\alpha\, \frac{dn_G}{dk^2_T\,d\alpha}\,
  \Theta(t- t_c )\ ,
  \nonumber
\end{equation}
 where   the number of radiated gluons as a function 
of $\alpha$ and $\vec k_T$ is \cite{guber}
\begin{equation}
\frac{dn_G}{d\alpha\,dk_T^2} =
\frac{4\alpha_s(k_T^2)}{3\,\pi}\;\frac{1}{\alpha\,k_T^2}
\nonumber
\end{equation}
\item  the time  dependence of the gluon radiation,
controlled by the parameter. 
$t_0 ={(m_N\,x_{Bj})}^{-1}=0.2 fm /x_{Bj}$
\end{itemize}

The final result for $n_G$ is ,
\begin{equation}
n_G(t) = \frac{16}{27}\,\left\{
{\rm ln}\left(\frac{Q}{\lambda}\right)\,+\,
{\rm ln}\left(\frac{t\,\Lambda_{QCD}}{2}
\right)\,{\rm ln}\left[\frac{{\rm ln}(Q/\Lambda_{QCD})}
{{\rm ln}(\lambda/\Lambda_{QCD}}\right]\right\}\ ,
\label{60}
 \end{equation}
for $t < t_0$, and 
 \begin{eqnarray}
n_G(t) &=& \frac{16}{27}\,\left\{
{\rm ln}\left(\frac{Q}{\lambda}
\,\frac{t_0}{t}\right)\,+\,
{\rm ln}\left(\frac{t\,\Lambda_{QCD}}{2}
\right)\,{\rm ln}\left[\frac{{\rm ln}(Q/\Lambda_{QCD}
\sqrt{t_0/t})}
{{\rm ln}(\lambda/\Lambda_{QCD})}\right]
\right.\nonumber\\ &+& \left.
{\rm ln}\left(\frac{Q^2\,t_0}{2\,\Lambda_{QCD}}
\right)\,{\rm ln}\left[\frac{{\rm ln}(Q/\Lambda_{QCD})}
 {{\rm ln}(Q/\Lambda_{QCD}\,\sqrt{t_0/t})}\right]\right\}
\ ,
\label{65}
\end{eqnarray}
for $t > t_0$, with saturation at  $t > t_0\,Q^2/\lambda^2 = 2\nu/\lambda^2$.
Using Eqs. (\ref{50}), (\ref{60}), and  (\ref{65}),  one obtains the effective debris-nucleon cross section
in the following form

\begin{equation}
\sigma_{eff}(t)=\sigma_{tot}^{NN}+
\sigma^{MN}_{tot}\Bigl[n_M(t) +
n_G(t)\Bigr]\ 
\label{70}
\end{equation}

\begin{figure}[hc]
\centerline{\includegraphics[height=6.0cm,width=6.0cm]{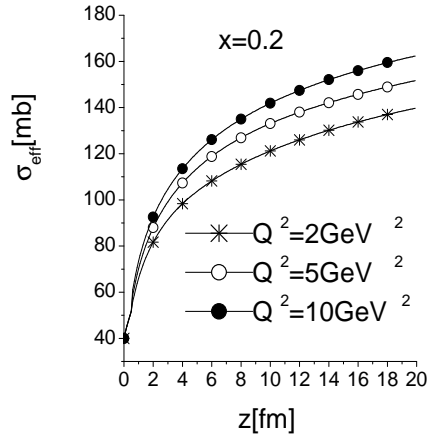}}
\vspace*{-0.2cm}
\caption[]{The debris-nucleon effective cross section (Eq. (\ref{70})) plotted {\it vs} \,\,the distance $z$ for a fixed 
value of the Bjorken scaling variable  $x \equiv x_{Bj}$ and various values of the four-momentum transfer 
 $Q^2$ (after  Ref. \cite{ck}).}
\label{fig2}
\end{figure}

The time (or $z$, for a particle moving with the speed of light) dependence of
$\sigma_{eff}(t)$ is shown in Fig. \ref{fig2} for a fixed value of $x_{Bj}$ and 
various values of $Q^2$.
 
\section{The Semi-Exclusive DIS Process A(e,e'B)X}
This process has been discussed in \cite{fs,cks} within the 
 Plane Wave Impulse Approximation (PWIA). If we consider, for ease of presentation,
 a deuteron $(D)$ target, the PWIA process consists of the hard interaction of $\gamma*$ 
 with a parton of,  e.g. the neutron, the creation of a debris, with the 
 spectator ($s$) proton recoiling and  detected  in coincidence
 with the scattered electron
  (cf. Fig. 1, left, disregarding the wavy lines that represent the FSI between the debris and the spectator nucleons).
 The PWIA cross sections reads as follows
 
 \begin{figure}[htb]
\centerline{\includegraphics[height=6.5cm,width=6.5cm]{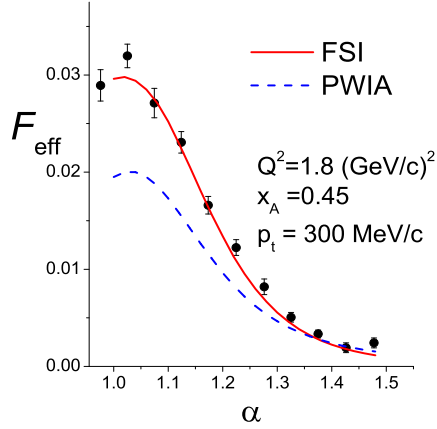}}
\caption[]{The effective  semi exclusive deuteron structure function (Eq. (\ref{feff})) calculated including (FSI) and 
omitting (PWIA) the FSI of the debris, compared with preliminary experimental data from Jlab \cite{kuhn}. 
$\alpha$ and  $p_t$ are the light cone fraction and perpendicular  momentum component of the detected proton and
$x_A=x_{Bj}/(2-\alpha)$ (after \cite{claleo}).}
\label{fig3}
\end{figure}
\begin{equation}
\frac{d\sigma}{dx dQ^2\ d\alpha dp_T^2}
=
K(x_{Bj},Q^2,p_s)\, n_D(|{\bf p}_s|) \, F_2^{N/D}(Q^2,x_{Bj},p_s),
\label{crossPWIA}
\end{equation}
is obtained, where $p_s =(p_{s}^0,{\bf p}_s)$ is the four-momentum of the
 recoiling detected nucleon 
, $\alpha = [p_{s}^0 - |{\bf p}_s|\cos \theta_s]/M$ ($\theta_s$ being the  
 nucleon emission
angle with respect to the direction of $\bf q$), $K(x_{Bj},Q^2,p_s)$ is 
 a kinematical factor,
$F_2^{N/A}(Q^2,x_{Bj},p_s) = 2x_{Bj} F_1^{N/A}(Q^2,x_{Bj},p_s)$ is  the DIS structure
function of  the  hit nucleon,
and $n_D$  the nucleon  momentum distribution, i. e.
\begin{equation}
n_D(|{\bf p}|)=\frac13\frac{1}{(2\pi)^3} \sum\limits_{{\cal
M}_D} \left |\int d^3 r 
 \Psi_{{1,\cal
M}_D}( {\bf r})\exp(-i{\bf p r}/2) \right|^2.
\label{dismom}
\end{equation}
where $\Psi_{{1,\cal
M}_D}( {\bf r})$ is the deuteron wave function. When the FSI of the debris is taken into account, the momentum distribution is
replaced by the distorted momentum distribution
\begin{equation}
 n_D^{FSI}( {\bf p}_s,{\bf q}) =
\frac13\frac{1}{(2\pi)^3} \sum\limits_{{\cal
M}_D} \left | \int\, d  {\bf r} \Psi_{{1,\cal
M}_D}( {\bf r}) S( {\bf r},{\bf q}) \chi_f^+\,\exp (-i
{\bf p}_s {\bf r}) \right |^2,
 \label{dismomfsi}
 \end{equation}
where $\chi_f$ is the  spin function of the spectator nucleon, ${\bf q}$ is the three-momentum transfer
(oriented along the $z$ axis),
and 
\begin{equation}
S( {\bf r},{\bf q}) = 1-\theta(z)\, \frac{\sigma_{eff}(z,Q^2,x)(1-i\alpha)}{4\pi b_0^2}\,
\exp(-b^2/2b_0^2)
\label{gama}
\end{equation}
 takes care of 
the  final state interaction
between the debris and the  spectator.
The approach can  readily be generalized to complex nuclei for
which the experiments are difficult to carry out. However, preliminary experimental data for the deuteron have
already
been obtained at Jlab \cite{kuhn}. A limited set of these data is shown in Fig. 3, where the quantity
\begin{equation}
F_{eff} \left( \frac{x_{Bj}}{2-\alpha},p_T,Q^2 \right)\equiv \frac{\sigma^{exp}}{K\cdot n_D(|{\bf p}|)}
\label{feff}
\end{equation}
is compared with theoretical calculations \cite{claleo} which omit (Eq. (\ref{dismom})) and include
(Eq. (\ref{dismomfsi})) the FSI of the debris; it can be seen that the latter is extremely important.


\section{Hadronization and grey track production}\label{grey}

The dominant channels of DIS  are the ones in which
the recoiling nucleus $B$  breaks apart to fragments; the investigation of the
nature and the  kinematical dependence of these fragments can   shed light on the time
evolution of the jet in the nuclear medium. This seems particularly true in the processes producing
 grey tracks,
 which are hadrons, predominantly protons, with momenta in the range  of few hundred MeV/c. The basic
mechanism of grey track production  is the inelastic interaction
of the jet with the spectator nucleons of the target, which recoil in the given momentum interval 
and are detected (cf. Fig. 1, right).
The $Q^2$ and $x_{Bj}$ dependence of the average number of 
grey tracks produced in the
Fermilab E665  experiment \cite{e665} (  protons with momentum 
$200-600\,\, MeV/c$ produced in  $\mu-Xe$ and $\mu-D$ DIS  
at $490\,\, GeV$ beam energy), have been recently analyzed \cite{grey} within the theoretical framework which employs 
the effective debris-nucleon cross section of \cite{ck}.
\begin{figure}[!hb]
\centerline{\includegraphics[height=5.8cm,width=6.0cm]{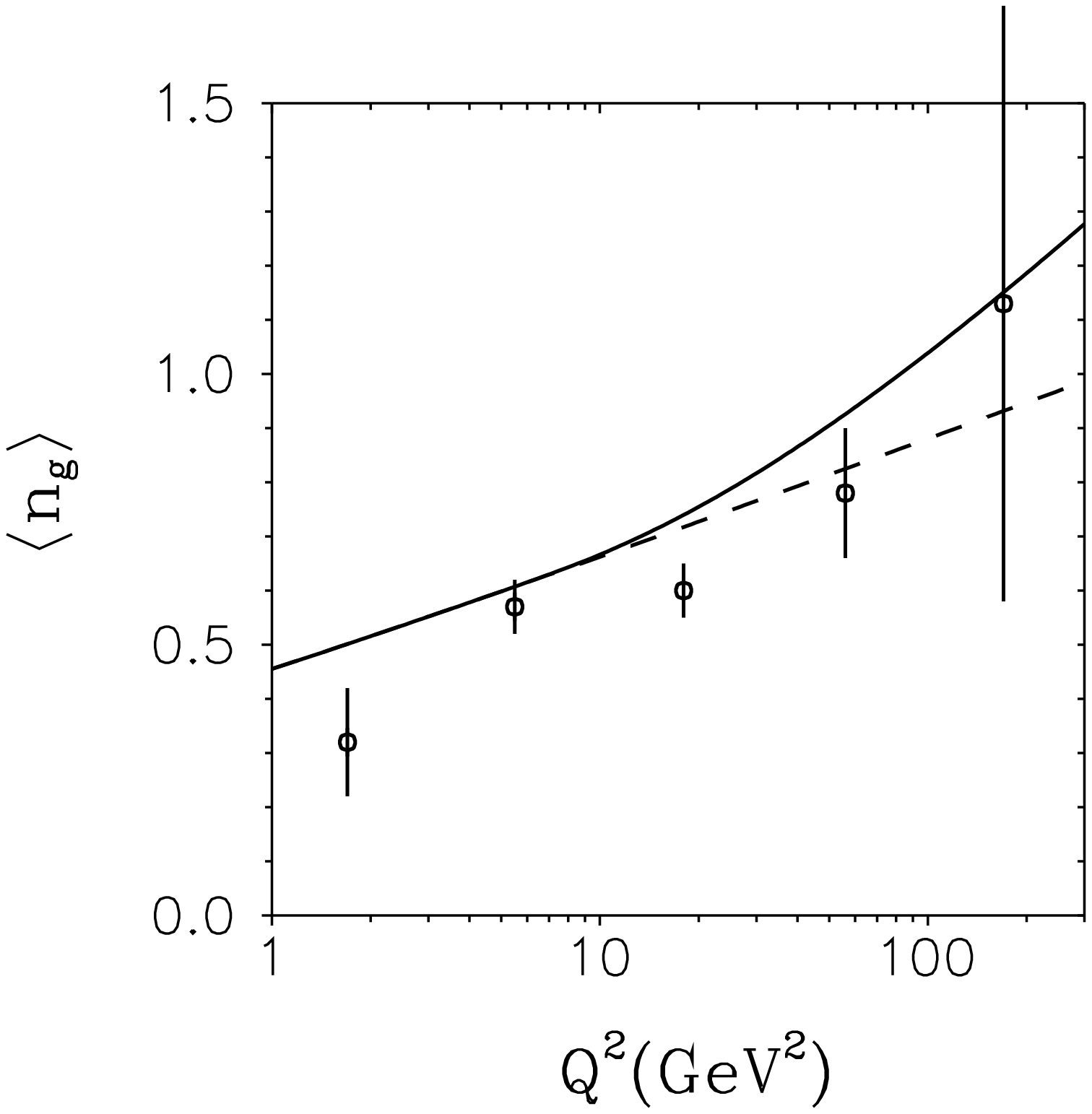}
\hspace{0.5cm}\includegraphics[height=5.8cm,width=5.5cm]{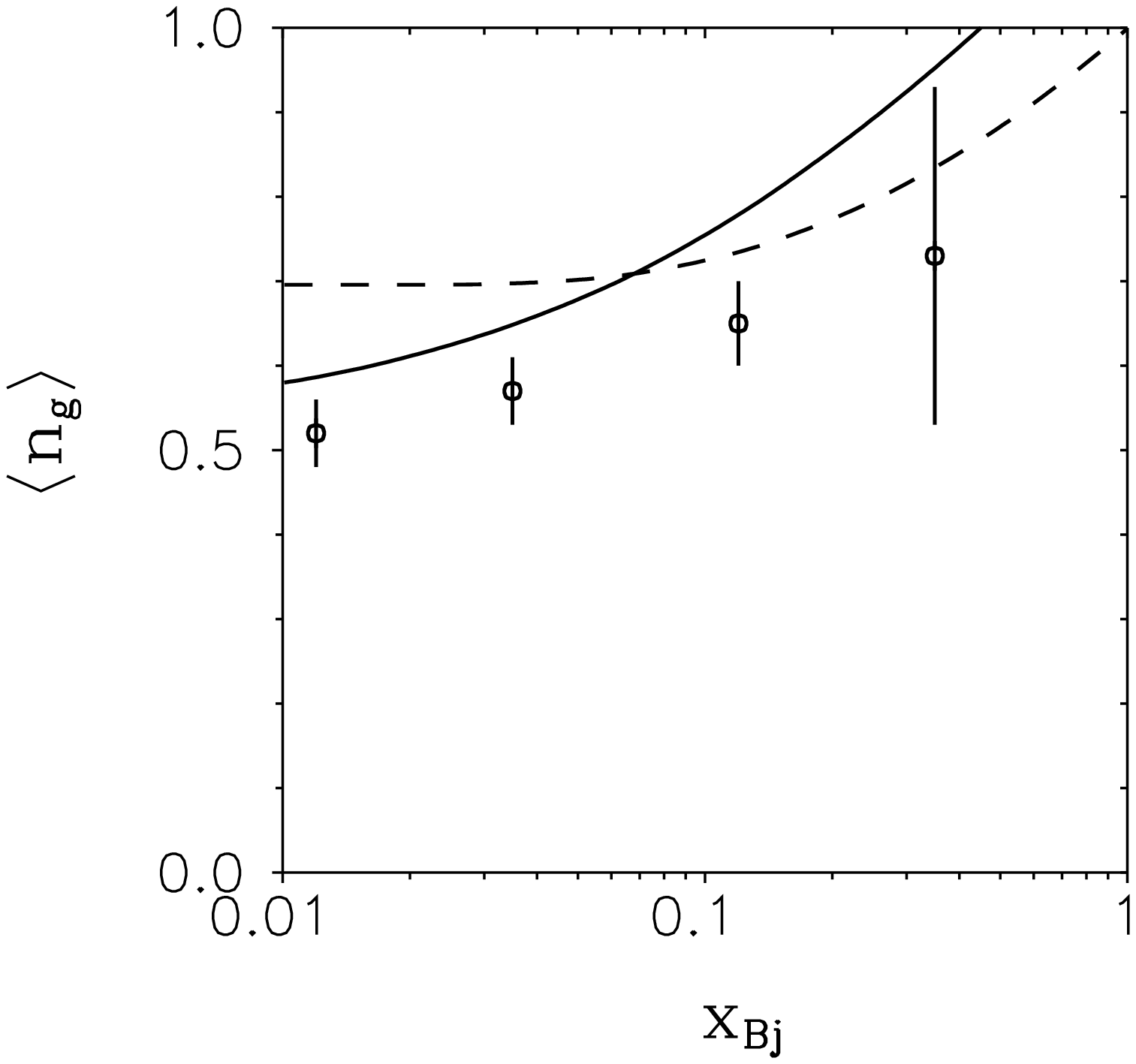}}
\caption[]{{\it Left}: mean number of grey tracks $< n_g >$ produced in the $\mu - Xe$ DIS experiment \cite{e665}
{\it vs}  $Q^2$ in the non-shadowing region with fixed $x_{Bj}=0.07$
 (dashed). {\it Right}:  mean number of grey tracks
 $< n_g >$  {\it vs}  $x_{Bj}$  with fixed value of
$Q^2=14.3\,\, GeV^2$ (dashed). In both Figures  the solid curve 
includes the $Q^2-x_{Bj}$ correlation found in the experiment
(after Ref. \cite{grey}).}
\label{fig4}
\end{figure}
The basic elements of the approach are the following ones:
\begin{itemize}
 \item  DIS  on a bound nucleon  occurs at coordinate $(\vec b,z)$ and 
   the debris
 propagates through the nucleus interacting with
spectator  nucleons via
$\sigma_{eff}(z-z')$. The  mean number of collisions 
 (plus 
 the recoiling nucleon formed in the hard $\gamma*-N$ act)  is
 \begin{equation}
{\langle \nu_c\rangle = \int d^2b\,
\int\limits_{-\infty}^\infty dz\,\rho_A(\vec b,z)
\int\limits_z^\infty dz'\,\rho_A(\vec b,z')\,
\sigma_{eff}(z-z')\,+\,1\ .}
\label{medio}
\end{equation}
 where $\rho_A$ is the nuclear density and $\sigma_{eff}(t)=\sigma_{in}^{MN}[n_M(t)+n_G(t)]$
with $\sigma_{in}^{MN}=\sigma_{tot}^{\pi\,N} -\sigma_{el}^{\pi\,N} - \sigma_{dif}^{\pi\,N}=17.7\,mb$. 
 \item the mean number of collisions has been calculated according to Eq. (\ref{medio}),
 and the mean number of grey tracks   $<n_g>$  has been obtained  using
the empirical relation found in \cite{e665}, {\it viz.}
 \begin{equation}
{<n_g> =\frac{\langle {\nu}_c\rangle - (2.08\pm0.13)}{(3.72\pm0.14)}\,}
\label{empirical}
\end{equation}
  \end{itemize} 

\noindent The results of the parameter-free calculations, which  are shown in Fig. 4, exhibit a good agreement
with the experimental data; worth being mentioned is the $Q^2$ dependence of the data, which is explained
by the adopted hadronization model, namely by the gluon radiation mechanism.
It turns out, therefore, that the observed   $Q^2$ dependence of the average number of grey tracks  is a
very sensitive tool to discriminate  between different models of hadronization.

\section{Conclusions}
The main conclusions of my talk can be summarized as follows:
\begin{itemize}
\item a time-dependent  debris-nucleon cross section has been obtained; 
it  incorporates 
both non perturbative (string) and $Q^2$-dependent perturbative (gluon radiation)
 effects; it accounts for 
the bulk of the FSI interaction of the debris created by the hard interaction of a bound 
nucleon with $\gamma*$; it should be used in all kinds of DIS off nuclei to describe the FSI
of the hit nucleon debris with the spectator nucleons;
\item  the (parameter-free) calculation of the Semi Exclusive DIS off the deuteron 
$D(e,e'p)X$
   exhibits good  agreement with
  preliminary data from Jlab; a systematic investigation of the $x_{Bj}$ and $Q^2$
  dependence of these processes at Jlab and HERA energies would shed further light on the 
   hadronization mechanism;
\item the (parameter-free) calculation of the $Q^2$ and $x_{Bj}$ dependence of 
 the average number of  
grey tracks produced 
 in  Deep Inelastic $\mu-Xe$ scattering exhibits  a very satisfactory agreement 
  with the experimental data, thanks to the gluon radiation
   mechanism 
  for hadron production.
  
\end{itemize}

\section*{Acknowledgments}I am grateful to the Organizers of the Workshop for the invitation and to 
 Leonid Kaptari and Boris Kopeliovich for a fruitful and stimulating collaboration.

\vfill\eject
\end{document}